\documentclass{ws-p8-50x6-00}

\makeindex

\begin{document}

\title{Scaling and duality in the superconducting phase transition}

\author{F.S. Nogueira\index{NOGUEIRA, F.S.}}

\address{Institut f\"ur Theoretische Physik, Freie Universit\"at Berlin, 
Arnimallee 14, 14195 Berlin, Germany\\ 
E-mail: nogueira@physik.fu-berlin.de}


\maketitle

\index{duality}

\abstracts{The field theoretical approach to duality in the 
superconducting phase transition is reviewed. Emphasis is given 
to the scaling behavior, and recent results are discussed.}

\section{Introduction}

The renewed interest for critical fluctuations in superconductors\index{superconductors} 
is due to the enormous variety of interesting phenomena observed in 
the last decade. In a zero external field regime, non-classical values 
of the critical exponents and amplitude ratios were measured\cite{NOGUKamal}. 
In the observed critical region we have $-0.03<\alpha<0$, 
$\nu\approx 0.67$ and $A_+/A_-\approx 1.065$. These values were measured 
for bulk samples of YBa$_{2}$Cu$_{3}$O$_{7-\delta}$ (YBCO). They 
indicate that the observed critical region corresponds to the 
$XY$ universality class. In the non-zero field regime, the thermal 
fluctuations make the vortex lattice melt. For high fields 
this transition is known to be of first-order\cite{NOGUExp}. 

Most thermal fluctuation effects 
in superconductors\index{superconductors} can be 
understood using the Ginzburg-Landau 
(GL) model\index{Ginzburg-Landau model}. In zero field, 
the GL model is given by 
\begin{equation}
\label{NOGUGL}
L=\frac{1}{2}(\nabla\times{\bf A})^2+|(\nabla-ie{\bf A})\phi|^2
+m^2|\phi|^2+\frac{u}{2}|\phi|^4.
\end{equation} 
The different regimes of the GL model are controlled by the size of 
the Ginzburg parameter $\kappa=\sqrt{u/2e^2}$. 
At a weak-coupling regime ($\kappa\ll 1$) 
the critical fluctuations in the GL model 
lead to a fluctuation induced first-order phase transition\cite{NOGUHLM}. 
This scenario no longer holds in a strong-coupling regime 
($\kappa\gg 1$), where a second-order phase transition 
takes place\cite{NOGUKleinTric,NOGUDasgupta}. Usually it is very difficult 
to access the strong-coupling regime of the GL model through 
conventional perturbative methods. An alternative approach uses 
duality arguments. Duality\index{duality} 
allows us to transform a strong-coupling 
problem into a weak-coupling one. A well-known example is the 2d Ising 
model where suach a transformation has an additional feature:  
Two-dimensionality makes the model self-dual, which leads   
to an exact determination of the critical temperature\cite{NOGUKramers}. In 
three dimensions life is more complicate, but duality remains a 
powerful tool. In the case of the GL model, lattice duality 
studies\cite{NOGUDasgupta} helped to conclude that the transition must be of 
second-order in the strong coupling regime. A deeper point of view  
was pioneered by Prof. Kleinert who developed a scaling 
(continuum) limit of the lattice dual model\cite{NOGUKleinert,NOGUKleinTric,NOGUKiometzis} which gives a field theoretic description of vortex lines. 
Using this field theoretical approach, Prof. Kleinert made the 
remarkable discovery that a tricritical point\index{tricritical point} exists in the phase 
diagram of a superconductor\index{superconductors}. 
The tricritical point\index{tricritical point}  
separates the first- and second-order phase transition regimes of the 
superconductor\index{superconductors}. We 
shall see in the next section that this discovery 
has far reaching consequences and is useful even in a nonzero field 
regime. 

In zero field, the vortex lines are closed loops, and the corresponding 
field theory features a {\it disorder parameter} field $\psi$ (as opposed 
to the {\it order parameter} field $\phi$). The field $\psi$ describes 
a grand canonical ensemble for vortex loops, and  
$|\psi|^2$ gives the vortex density. The duality transformation 
has transformed a  
field theory, where the basic objects are the Cooper pairs, into another  
one, where the basic objects are vortex lines. In the case of 
the GL model, currents interact through the eletromagnetic vector 
potential ${\bf A}$, while in the disorder field theory they 
interact through a fluctuating fielf which is proportional to the  
{\it magnetic induction} field ${\bf h}$. The 
simplest example of duality is obtained in the London 
limit, where the amplitude fluctuations are frozen. 
There, the dual Lagrangian corresponds to the London model: 
\begin{equation}
L_{\rm London}^{\rm dual}
=\frac{1}{2}(\nabla\times{\bf h})^2+\frac{m_A^2}{2}{\bf h}^2 
+im_A{\bf J}_v\cdot{\bf h},
\end{equation}
where $m_A$ is the photon mass and ${\bf J}_v$ is the vortex current. 
The full disorder field theory corresponds to a generalization of this 
model. Intuitively this can be done as follows: in the classical 
limit, thermal fluctuations are absent and, therefore, there are no vortex 
loops. The only way to create vortices is by applying an 
external magnetic field. This creates vortex lines 
parallel to this field but no loops. The 
classical solution is well known in this case and corresponds to the 
Abrikosov vortex lattice\cite{NOGUAbrikosov}. In the context of the  
model (2), 
the external magnetic field couples linearly to the induction 
field ${\bf h}$. Thermal fluctuations create additional vortex loops. 
They are closed as a consequence of Ampere's law which gives 
$\nabla\cdot{\bf J}_v=0$. In the disorder field theory of 
fluctuating vortex loops, the coupling ${\bf J}_v\cdot{\bf h}$ in 
(2) becomes a minimal coupling, and the result is Kleinert's dual 
model \cite{NOGUKleinert,NOGUKiometzis}

\begin{equation}
\label{NOGUdual}
L_{d}=\frac{1}{2}[(\nabla\times{\bf h})^2+m_A^2{\bf h}^2]
+|(\nabla-ie_d{\bf h})\psi|^2
+m_{\psi}^2|\psi|^2+\frac{u_{\psi}}{2}|\psi|^4, 
\end{equation}                
where $e_d=2\pi m_A/e$ is the dual charge. 

In the next sections we shall review the recent results on Kleinert's 
model.  
We shall discuss the relation between  
the tricritical point\index{tricritical point} as obtained 
from dual model 
and the tricritical point\index{tricritical point} in the original GL 
model. It will be shown that the tricritical point\index{tricritical point} in the GL model is 
of a Lifshitz type\cite{NOGUSelke}. 
In the GL model, the most important manifestation of the 
Lifshitz point\index{Lifshitz point} 
is a negative sign of the anomalous 
dimension of the order parameter. 
Duality\index{duality} transforms the tricritical\index{tricritical point} Lifshitz point\index{Lifshitz point} into an 
ordinary tricritical point\index{tricritical point}. As a consequence, the sign of the 
anomalous dimension of the disorder parameter will be positive. 

In Section 3 we discuss the scaling behavior of the dual model,  
which allows for many possibilities of scaling due to 
the massiveness of the induction field\cite{NOGUdeCalan}. Of course, only 
one scaling corresponds to the superconducting phase transiton 
as obtained from the original GL model, that is, 
the transition which is governed by an infrared stable charged 
fixed point. The other scalings correspond to crossover regimes. 
In principle, all 
these regimes can be observed experimentally. Ironically, 
the true superconducting phase transition remains the most difficult 
to be observed experimentally. In fact, the critical region of 
this charge fluctuation regime is very small\cite{NOGUKleinert}. 
The regime very often probed is the XY regime\cite{NOGUKamal} and 
we will see that there is a corresponding scaling in the dual 
Lagrangian that corresponds to it. A scaling that deserves 
some experimental attention is the Kleinert scaling\cite{NOGUKiometzis} 
which is characterized by the {\it exact} value $\nu'=1/2$ 
of penetration depth exponent. 
This scaling seems to be verified in two experiments using 
thin films of YBCO at optimal doping\cite{NOGUPaget,NOGULin}. At optimal 
doping 3d fluctuations are still dominant even in thin films of 
YBCO and a three-dimensional model is still relevant.

\section{Tricritical \index{tricritical point} 
and Lifshitz Point\index{Lifshitz point}}

Let us briefely review Kleinert's discovery of the tricritical point\index{tricritical point}. 
In order to make the discussion simple, we will take 
advantage of the intuitive point of view adopted in the 
introduction. The technical details can be found in the textbook 
of Kleinert\cite{NOGUKleinert} and in the seminal paper Ref.\cite{NOGUKleinTric}. 

Let us integrate out the induction field ${\bf h}$ in Eq. (\ref{NOGUdual}). 
We obtain the following effective action: 
\begin{eqnarray}
\label{NOGUdualMF}
S_{\rm eff}&=&\frac{1}{2}{\rm Tr}\,\ln[(-\partial^2+m_A^2+e^2_d |\psi|^2)\delta_{\mu\nu}
+(1-1/a)\partial_\mu \partial_\nu]\nonumber\\
&+&\frac{e_d^2}{2}\int d^3 r\int d^3 r'
j_\mu({\bf r})D_{\mu\nu}({\bf r},{\bf r}')j_\nu({\bf r}')\nonumber\\
&+&\int d^3 r \left(|\nabla\psi|^2+m_\psi^2 |\psi|^2+\frac{u_\psi}{2}
|\psi|^4\right),
\end{eqnarray}       
where $j_\mu$ is the $\mu$ component of the current operator 
${\bf j}=ie_d(\psi^\dag \nabla\psi-\psi\nabla\psi^\dag)$ and 
the limit $a\to 0$ must be taken at the end in order to inforce 
the constraint $\nabla\cdot{\bf h}=0$. The kernel 
$D_{\mu\nu}(r,r')$ is the inverse of the operator inside the 
$Tr\ln$. Now we will perform a Landau expansion of the effective action 
(\ref{NOGUdualMF}). As usual, in this expansion we assume that $\psi$ has no 
spatial variation. In this way we obtain the following free energy 
density: 
\begin{eqnarray}
\label{NOGUFE}
{\cal{F}}&=&[m_\psi^2+e_d^2 D_{0;\mu\mu}(0)]|\psi|^2+\frac{1}{2}
\left[u_\psi-e_d^4 \int\frac{d^3 k}{(2\pi)^3}
\hat{D}_{0;\mu\nu}(k)\hat{D}_{0;\nu\mu}(k)
\right]|\psi|^4\nonumber\\
&+&\frac{e_d^3}{3}\int\frac{d^3 k}{(2\pi)^3}
\hat{D}_{0;\mu\lambda}(k)\hat{D}_{0;\lambda\delta}(k)
\hat{D}_{0;\delta\mu}(k)|\psi|^6, 
\end{eqnarray}
where $D_{0;\mu\nu}({\bf r}-{\bf r}')$ is the kernel 
$D_{\mu\nu}({\bf r},{\bf r}')$ for 
$\psi=0$ and $\hat{D}_{0;\mu\nu}(k)$ is its Fourier transform 
which is given by
\begin{equation}
\hat{D}_{0;\mu\nu}(k)=\frac{1}{k^2+m_A^2}\left(\delta_{\mu\nu}
-\frac{k_\mu k_\nu}{k^2}\right).
\end{equation}
After calculating the integral in the $|\psi|^4$ term in 
(\ref{NOGUFE}), we see that it will be negative if 
$u_\psi<4\pi^3 m_A^3/e^4$. When this happens, we have a first-order 
phase transition scenario. Thus, it is clear that the point 
$m^2_\psi =2\pi m_A^3/e^2$, $u_\psi=4\pi^3 m_A^3/e^4$ corresponds 
to a tricritical point\index{tricritical point} (we have redefined $m_\psi^2$ by absorbing 
a factor $e_d^2 \Lambda/\pi^2$, where $\Lambda$ is the ultraviolet 
cutoff).

Now we can ask the following question: How does the tricritical point\index{tricritical point} 
manifest itself in the GL model? From a RG point of view we have the  
following scenario:  
let us define the renormalized dimensionless couplings 
$f=e_r^2/\mu$ and $g=u_r/\mu$, where $e_r$ and $u_r$ are the 
renormalized couplings and $\mu$ is a running scale. The fixed 
point structure in the $gf$-plane contains 
four fixed points\cite{NOGUBerg,NOGUFolk,NOGUHerbut,NOGUdeCalan1}.  
Two of them are uncharged: the Gaussian fixed point corresponding 
to mean-field behavior of an uncharged superfluid, and the $XY$ fixed 
point that governs the critical behavior of $^4$He superfluid. The 
other two fixed points are charged:  
the infrared stable fixed point that governs the critical behavior 
of the superconductor\index{superconductors} and 
the tricritical fixed point\index{tricritical point}, which is 
infrared stable along a line connecting the Gaussian and the 
tricritical fixed point\index{tricritical point}, being unstable in the $g$-direction. The line 
connecting the Gaussian fixed point and the tricritical point\index{tricritical point} is called 
the tricritical line. This line separates the regimes of first- and 
second-order phase transition. The schematic flow diagram is shown in 
Fig.1. 
\begin{figure}
\vspace{-1.5cm}
\centerline{\psfig{figure=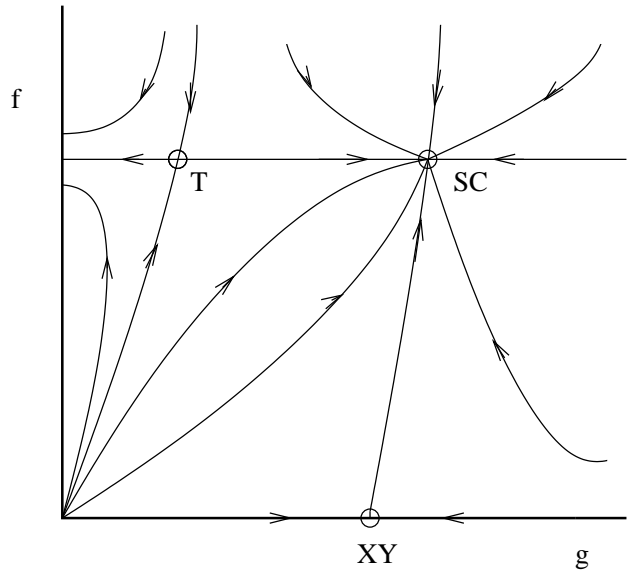,height=15truecm,angle=-90}}
\vspace{-1.5cm}
\caption{Flow diagram for the GL model. The labels $T$ and $SC$ are 
for tricritical and superconducting, respectively.}
\end{figure}
The tricritical point\index{tricritical point} obtained in 
Kleinert's duality map\cite{NOGUKleinTric,NOGUKleinert} corresponds to 
the tricritical fixed point\index{tricritical point} of the RG picture. However, this tricritical 
point\index{tricritical point} obtained directly in the GL model is of a different nature 
due to the local gauge symmetry. Indeed, we have suggested 
recently\cite{NOGUNogueira} 
that the tricritical point\index{tricritical point} of the GL model is of 
a Lifshitz type. This picture is founded on the behavior of the 
2-point bare correlation function, which is given at 1-loop and 
$d=3$ by
\begin{equation}
\label{NOGU2pfGL}
\tilde{W}^{(2)}(p)=\frac{1}{p^2+m^2+\Sigma(p)},
\end{equation}
with the self-energy
\begin{equation}
\label{NOGUSE}
\Sigma(p)=-\frac{m}{2\pi}(u+e^2)-
\frac{e^2}{4\pi|p|}(p^2-m^2)
\left[\frac{\pi}{2}+
\arctan\left(\frac{p^2-m^2}{2m|p|}\right)
\right].
\end{equation}
In writing the above equations, we have absorbed in the bare mass
a contribution with a linear dependence on the ultraviolet cutoff 
$\Lambda$. From Eq. (\ref{NOGU2pfGL}) we see that  
$\tilde{W}^{(2)}(p)$ has a real pole in  the critical regime ($m^2=0$) 
at a nonzero momentum, besides 
the usual pole at $p=0$. In the above 1-loop calculation 
this pole is at $|p|=e^2/4$. This means that $\tilde{W}^{(2}(p)<0$ 
when $|p|<e^2/4$ and the bare 2-point correlation function 
violates the infrared bound $0<\tilde{W}^{(2)}(p)\leq 1/p^2$, usually 
satisfied for pure scalar models\cite{NOGUSpencer}. This violation of the 
infrared bound explains the negative sign of the $\eta$ exponent 
usually found in RG calculations\cite{NOGUBerg,NOGUFolk,NOGUHerbut,NOGUdeCalan1}. 
The negativeness of the $\eta$ exponent is also confirmed by 
recent Monte Carlo calculations\cite{NOGUNguyen}. Since the 
2-point critical correlation function changes its sign, it follows 
that the same sign change happens in the 1-particle irreducible 
2-point function $\tilde{\Gamma}^{(2)}(p)$, which is the 
coefficient of the quadratic term in the effective action $\Gamma$. 
This change of sign with momentum at the critical point is 
a behavior characteristic of a Lifshitz point\index{Lifshitz point}\cite{NOGUSelke}. 
It is worth to mention that in scalar models of 
Lifshitz points\index{Lifshitz point} the sign of $\eta$ is also negative for 
dimension $d=d_c-1$ where $d_c$ is the corresponding critical dimension. 
For instance, a fixed dimension calculation in a $1/N$ 
expansion gives, for the isotropic Lifshitz point\index{Lifshitz point} in $d=7$ ($d_c=8$ in 
this case), $\eta_{l4}\approx -0.08/N$~\cite{NOGUHornreich1}.    

The phase transition scenario that emerges is the following. The 
phase diagram in the $\kappa^2-T$ plane contains three phases: 
the normal phase, the type I and the type II regime. The type I 
regime is separated from the type II regime by a line terminating 
at a tricritical\index{tricritical point} Lifshitz point\index{Lifshitz point}, the latter belonging to a line that 
separates the normal phase from the two other phases. The phase 
diagram is therefore quite similar to that of the so-called 
$R-S$ model\cite{NOGURedner}. In the $R-S$ model the phase diagram is drawn 
in the $X-T$ plane where $X=S/R$, $S$ and $R$ being the couplings 
of the model. The three phases of the $R-S$ model are 
paramagnetic, ferromagnetic and helical. Thus, the $R-S$ model 
differs from ordinary magnets by the presence of a modulated regime 
for the order parameter, the helical phase. If  
$\kappa^2$ plays a role analogous to $X$, we see that the type II 
regime is analogous to the helical phase. Indeed, the type II regime 
should correspond to a modulated order parameter, as can be seen 
experimentally by applying a magnetic field, which leads to the 
formation of the Abrikosov vortex lattice. The type I regime,  
on the other hand, must 
be associated to the ferromagnetic phase since it corresponds to a 
uniform order parameter.

\section{Kleinert's\index{KLEINERT, H.} Scaling in the Dual Model}

Let us discuss the renormalization of the dual model Eq. (\ref{NOGUdual}). 
An important feature of the dual model is the presence of the two 
mass scales $m_\psi$ and $m_A$. This fact allows some freedom in 
the scaling of the model which is clarified in Ref.\cite{NOGUdeCalan}. 

The scaling is defined by the behavior near $T_c$ of the {\it bare} 
ratio 
\begin{equation}
\label{NOGUkappad}
\kappa_d^2=\frac{m_\psi^2}{m_A^2}.
\end{equation}
It must be observed that $m_\psi^2\sim t$, where $t$ is the reduced 
temperature. Since the following argument is valid to all 
orders, we are assuming that the critical temperature contains 
already all the fluctuations. Thus, in the bare mass we are using 
a {\it renormalized} critical temperature.  
\footnote{For instance, at 1-loop the critical temperature would 
be corrected by a term proportional to the ultraviolet cutoff.} 
If we look to the scaling of the bare photon mass in the GL model, 
we see that we have also $m_A^2\sim t$. This is the main motivation 
of what we will call Kleinert's scaling\cite{NOGUKiometzis},   
where $\kappa_d$ is constant, just like the 
Ginzburg constant $\kappa$ in the GL model. 

We define the renormalized fields as $\psi_r=Z_\psi^{-1/2}\psi$ and 
${\bf h}_r=Z_h^{-1/2}{\bf h}$. From the Ward identities we conclude 
that the term $m_A^2{\bf h}^2/2$ does not renormalize. Thus, we 
obtain $m_{A,r}^2=Z_h m_A^2$. The renormalization of the 
remaining parameters is given by 
$m_{\psi,r}^2=Z_m^{-1}Z_\psi m_\psi^2$, $u_{\psi,r}=Z_{u_\psi}^{-1}Z_\psi^2 u$ 
and $e_{d,r}^2=Z_h e_d^2$. We observe from the renormalization of 
$e_d^2$ that the charge $e$ is not renormalized in the dual model. 
Let us introduce the dimensionless renormalized couplings 
$f_d=e_d^2/m_\psi$ and $g_\psi=u_{\psi,r}$. 
In order to obtain the flow equations we need to differentiate the 
renormalized quantities with respect $m_{\psi,r}$ keeping the bare 
quantities that {\it do not depend on} $t$ fixed. The following 
flow equations are easily obtained:
\begin{equation}
\label{NOGUflowmA}
m_{\psi,r}\frac{\partial m_{A,r}^2}{\partial m_{\psi,r}}=
(\eta_h+2+\eta_m-\eta_\psi)m_{A,r}^2,
\end{equation}
\begin{equation}
\label{NOGUflowfd}
m_{\psi,r}\frac{\partial f_d}{\partial m_{\psi,r}}=
(\eta_h+1+\eta_m-\eta_\psi)f_d,
\end{equation}
where the RG functions $\eta_h$, $\eta_m$ and $\eta_\psi$ are defined 
by
\begin{equation}
\eta_h=m_{\psi,r}\frac{\partial\ln Z_h}{\partial m_{\psi,r}},
\end{equation}
\begin{equation}
\eta_m=m_{\psi,r}\frac{\partial\ln Z_m}{\partial m_{\psi,r}},
\end{equation}
\begin{equation}
\eta_\psi=m_{\psi,r}\frac{\partial\ln Z_\psi}{\partial m_{\psi,r}}.
\end{equation} 
It is straightforward to see that the infrared stable fixed point 
corresponds to $f_d^*=0$. Therefore, the correlation length 
exponent is simply given by $\nu\approx 0.67$.  
Near the infrared stable fixed point, Eq. (\ref{NOGUflowmA}) becomes 
\begin{equation}
\label{NOGUlinearized}
m_{\psi,r}\frac{\partial m_{A,r}^2}{\partial m_{\psi,r}}\approx
\frac{1}{\nu}m_{A,r}^2. 
\end{equation}
This means that $m_{A,r}^2\sim m_{\psi,r}^{1/\nu}$. 
Since $m_{A,r}=\lambda^{-1}$ we have the {\it exact} penetration 
depth exponent $\nu'=1/2$. Thus, in Kleinert's scaling the exponent 
$\nu'$ remains classical to {\it all orders}.    
    
Other scalings corresponding to different physical 
situations are also possible in the dual model. For example, 
the scaling that gives the $XY$ universality class corresponds 
to taking $m_A$ {\it fixed}, that is, with no dependence on 
$t$. In this scaling $\kappa_{d,r}\to 0$ as $t\to 0$ an the 
penetration depth exponent is given by $\nu'=\nu/2$. This 
scaling is well known experimentally\cite{NOGUKamal}. 

Experimentally, Kleinert's scaling seems to have been probed by 
the authors of Ref.\cite{NOGULin} and more recently in  
Ref.\cite{NOGUPaget}. Both experiments used YBCO thin films at 
{\it optimal doping}. At optimal doping, 3d fluctuations are 
still more relevant, even in YBCO thin films, due to the 
strong coupling between the CuO planes. 

\section{Conclusion}          

Let us summarize the main lessons of this paper. First, the charged 
fixed point corresponds to a strong-coupling 
problem which is very difficult to be studied directly in the 
GL model where special perturbative or non-perturbative 
techniques must be employed. Duality\index{duality} 
gives in this case a powerful 
access towards the understanding of this problem since 
it provides a weak-coupling realization of the strong-coupling 
limit of the GL model. 
Second, the tricritical point\index{tricritical point} of the superconductor\index{superconductors} is a 
Lifshitz point\index{Lifshitz point}, and that is the reason why the anomalous 
dimension of the superconductor\index{superconductors} is negative.

An interesting perspective is the use of duality in a nonzero 
field problem. This approach is considerably more difficult since 
the phase diagram is richer. However, it can be hoped that the 
duality approach would be also helpful in this context.    

\section*{Acknowledgments}

The author would like to thank A. Pelster for his comments and 
suggestions. He acknowledges the Alexander von Humboldt 
foundation for the financial support.


\begin{thebibliography}{99}

\bibitem{NOGUKamal}  M.B. Salamon\index{SALAMON, M.B.}, 
J. Shi\index{SHI, J.}, N. Overend\index{OVEREND, N.}, and M.A. Howson\index{HOWSON, M.A.},  
{\it Phys. Rev. B} {\bf 47}, 5520 (1993); 
N. Overend\index{OVEREND, N.}, 
M.A. Howson\index{HOWSON, M.A.}, and I.D. Lawrie\index{LAWRIE, I.D.}, {\it Phys. Rev. Lett.} 
{\bf 72}, 3238 (1994); S. Kamal\index{KAMAL, S.},  
D.A. Bonn\index{BONN, D.A.},  
N. Goldenfeld\index{GOLDENFELD, N.}, 
P.J. Hirschfeld\index{HIRSCHFELD, P.J.}, 
R. Liang\index{LIANG, R.}, and  
W.N. Hardy\index{HARDY, W.N.}, 
{\it Phys. Rev. Lett.} {\bf 73}, 1845 (1994).

\bibitem{NOGUExp} E. Zeldov\index{ZELDOV, E.}, E. Maier\index{MAIER, E.}, 
M. Konczykowski\index{KONCZYKOWSKI, M.}, V.B. Geshkenbein\index{GESHKENBEIN, V.B.}, 
V.M. Vinokur\index{VINOKUR, V.M.}, and H. Shtrikman\index{SHTRIKMAN, H.},   
{\it Nature} (London) {\bf 375}, 373 (1995); 
M. Roulin\index{ROULIN, M.}, A. Junod\index{JUNOD, A.}, 
and E. Walker\index{WALKER, E.}, 
{\it Science} {\bf 273}, 1210 (1996); for a review 
see G. Blatter\index{BLATTER, G.}, M.V. Feigel'man\index{FEIGEL'MAN, M.V.}, 
V.B. Geshkenbein\index{GESHKENBEIN, V.B.}, 
A.I. Larkin\index{LARKIN, A.I.}, and V.M. Vinokur\index{VINOKUR, V.M.},   
{\it Rev. Mod. Phys.} {\bf 66}, 1125 (1994). 

\bibitem{NOGUHLM} B.I. Halperin\index{HALPERIN, B.I.}, 
T.C. Lubensky\index{LUBENSKY, T.C.}, and S.-K. Ma\index{MA, S.-K.}, 
{\it Phys. Rev. Lett.} {\bf 32}, 292 (1974); 
J.-H. Chen\index{CHEN, J.-H.}, T.C. Lubensky\index{LUBENSKY, T.C.}, 
and D.R. Nelson\index{NELSON, D.R.}, 
{\it Phys. Rev. B} {\bf 17}, 4274 (1978).

\bibitem{NOGUKleinTric} H. Kleinert\index{KLEINERT, H.}, 
{\it Lett. Nuovo Cimento} {\bf 35}, 405 (1982).

\bibitem{NOGUDasgupta} C. Dasgupta\index{DASGUPTA, C.} and 
B.I. Halperin\index{HALPERIN, B.I.},  
{\it Phys. Rev. Lett.} {\bf 47}, 1556 (1981).

\bibitem{NOGUKramers} H.A. Kramers\index{KRAMERS, H.A.} 
and G.H. Wannier\index{WANNIER, G.H.}, {\it Phys. Rev.} 
{\bf 60}, 252 (1941).

\bibitem{NOGUKleinert} H. Kleinert\index{KLEINERT, H.}, 
{\em Gauge Fields in Condensed 
Matter}, 2 vols. (World Scientific, Singapore, 1989).

\bibitem{NOGUKiometzis} M. Kiometzis\index{KIOMETZIS, M.}, 
H. Kleinert\index{KLEINERT, H.}, and A.M.J. 
Schakel\index{SCHAKEL, A.M.J.}, 
{\it Phys. Rev. Lett.} {\bf 73}, 1975 (1994); 
{\it Fortschr. Phys.} {\bf 43}, 697 (1995) and references therein.

\bibitem{NOGUAbrikosov} A.A. Abrikosov\index{ABRIKOSOV, A.A.}, 
{\it Soviet Phys. (JETP)} {\bf 5}, 1174 (1957).

\bibitem{NOGUTesanovic} Z. Te\u{s}anovi\'c\index{TE\u{S}ANOVI\'C, Z.}, 
{\it Phys. Rev. B} {\bf 59}, 6449 (1999).

\bibitem{NOGUNogueira} F.S. Nogueira\index{NOGUEIRA, F.S.}, 
{\it Phys. Rev. B} {\bf 62}, 14559 (2000).

\bibitem{NOGUSelke} W. Selke\index{SELKE, W.}, 
in {\it Phase Transitions and Critical 
Phenomena}, Vol. 15, eds. C. Domb\index{DOMB, C.} and 
J.L. Lebowitz\index{LEBOWITZ, J.L.} (Academic Press, London, 
1992), Vol. 15, pages 1-72, and references therein.

\bibitem{NOGUdeCalan} C. de Calan\index{DE CALAN, C.} and 
F.S. Nogueira\index{NOGUEIRA, F.S.}, 
{\it Phys. Rev. B} {\bf 60}, 4255 (1999). 

\bibitem{NOGUPaget} K.M. Paget\index{PAGET, K.M.}, 
B.R. Boyce\index{BOYCE, B.R.}, and T.R. Lemberger\index{LEMBERGER, T.R.}, 
{\it Phys. Rev. B} {\bf 59}, 6545 (1999).

\bibitem{NOGULin} Z.H. Lin\index{LIN, Z.H.} et al., 
{\it Europhys. Lett.} {\bf 32}, 573 (1995).

\bibitem{NOGUBerg} B. Bergerhoff\index{BERGERHOFF, B.} et al., 
{\it Phys. Rev. B} {\bf 53}, 5734 (1996).

\bibitem{NOGUFolk} R. Folk\index{FOLK, H.} and Y. Holovatch\index{HOLOVATCH, Y.}, 
{\it J. Phys. A} {\bf 29}, 3409 (1996).

\bibitem{NOGUHerbut} I.F. Herbut\index{HERBUT, I.F.} and 
Z. Te\u{s}anovi\'c\index{TE\u{S}ANOVI\'C, Z.},  
{\it Phys. Rev. Lett.} {\bf 76}, 4588 (1996); 
I.D. Lawrie\index{LAWRIE, I.D.}, {\it ibid.}, {\bf 78}, 979 (1997); 
I.F. Herbut\index{HERBUT, I.F.} and Z. Te\u{s}anovi\'c\index{TE\u{S}ANOVI\'C, Z.}, 
{\it ibid.} {\bf 78}, 980 (1997).

\bibitem{NOGUdeCalan1} C. de Calan\index{DE CALAN, C.},
A.P.C. Malbouisson\index{MALBOUISSON, A.P.C.}, 
F.S. Nogueira\index{NOGUEIRA, F.S.}, and 
N.F. Svaiter\index{SVAITER, N.F.}, {\it Phys. Rev. B} {\bf 59}, 554 (1999).

\bibitem{NOGUSpencer} J. Fr\"ohlich\index{FR\"OHLICH, J.}, B. Simon\index{SIMON, B.}, 
and T. Spencer\index{SPENCER, T.}, 
{\it Commun. Math. Phys.} {\bf 50}, 79 (1976).

\bibitem{NOGUNguyen} A.K. Nguyen\index{NGUYEN, A.K.} 
and A. Sudb\o\index{SUDB\O, A.}, {\it Phys. Rev. B} {\bf 60}, 
15307 (1999).

\bibitem{NOGUHornreich1} R.M. Hornreich\index{HORNREICH, R.M.}, 
M. Luban\index{LUBAN, M.}, and S. Shtrikman\index{SHTRIKMAN, S.}, 
{\it Phys. Lett. A} {\bf 55}, 269 (1975).

\bibitem{NOGURedner} S. Redner\index{REDNER, S.} and 
H.E. Stanley\index{STANLEY, H.E.}, {\it Phys. Rev. B} {\bf 16}, 
4901 (1977); {\it J. Phys. C} {\bf 10}, 4765 (1977).

\end{thebibliography}
\end{document}